%% file: euro.tex
\documentstyle{europhys}
\input euromacr

\input{epsf}

\def\pa{\partial}
 \def\G{\Gamma}
\def\o{\omega} 
\def\d{\delta} \def\D{\Delta}
\def\p{\phi}	\def\pn{\phi^0_n}
\def\r{\rho_n}
\def\b{\beta}

\begin{document}
\Date{}
\euro{}{}{}{}
\shorttitle{M. IGNATIEV {\it et al.} RESPONSE NEAR GLASS TRANSITION ETC.}
\title{Frequency-Dependent Response near the Glass Transition: A Theoretical Model}
\author{M. Ignatiev and Bulbul Chakraborty}
\institute{
The Martin Fisher School of Physics\\
Brandeis University\\
Waltham, MA 02254, USA}
\rec{}{}
\pacs{
\Pacs{64}{70Pf}{}
\Pacs{64}{60My}{}
\Pacs{82}{20Mj}{}
}
\maketitle
\begin{abstract}
We propose a simple dynamical model for a glass transition. The dynamics is
described by a  Langevin equation in a piecewise parabolic free energy landscape,
 modulated by a temperature dependent overall curvature. The zero-curvature
point marks a transition to a phase with broken ergodicity which we identify as
the glass transition.  Our analysis shows 
a connection 
between the high and low frequency response of systems approaching this 
transition.  

\end{abstract}

The glass transition in both spin systems and supercooled liquids is heralded by 
anomalously slow relaxations\cite{MCT,REVIEW}. Mode-coupling theory provides  a framework for
analyzing the dynamics of the supercooled state\cite{MCT} and has been
remarkably successful in predicting the general features of relaxations in a
highly viscous supercooled liquid.
However, detailed comparisons with experiments indicate that a complete
description of the glass transition is still lacking.  An alternative
description of glass formation relies on the presence of multi-valleyed free energy
surfaces\cite{Stillinger} and the activated dynamics resulting from the presence 
of these traps\cite{BOUCHAUD}.  

Recent experiments indicate that the approach to the glass transition has
some universal  features\cite{NAGEL} when viewed in terms of the
frequency-dependent  response of the system. In 
both  supercooled liquids\cite{NAGEL} and spin glasses\cite{NAGELSG}, a
frequency-independent behavior at high frequencies  and the presence of three 
distinct regimes
in the relaxation spectrum seem to be generic features of the approach to the 
glass transition. This behavior is different from 
that  observed near a
critical point where the high-frequency response is thought to be 
uninteresting\cite{Goldenfeld}. In this paper, we present a dynamical model which
provides a good description of the 
frequency-dependent response near the glass transition.


Dynamics of  systems approaching the glass transition have been modeled by
random walks  in environments of traps \cite{BOUCHAUD}. What distinguishes 
our model is the presence of an overall curvature modulating the landscape of
the traps, and the description of the dynamics $within$ the traps.
The curvature is introduced to model the coupling between a
set of  variables which are inherently frustrated and
another which tends to remove the degeneracy leading to the possibility of a
unique global minimum and many local minima of the free
energy\cite{LEIGU}.  
An example of
such variables could be the spins and strain fields in a spin
glass\cite{KARDAR} or a  frustrated antiferromagnet\cite{LEIGU},  or a local
orientational order parameter in a liquid coupled to local distortions of the
liquid\cite{DATTA}. The vanishing of the overall curvature is identified in our
model with the glass transition. This is reminiscent of
models where the glass transition is associated with an instability\cite{KLEIN}
or an avoided critical point\cite{KIVELSON}.  

The presence of the curvature ($R$) ensures that an equilibrium distribution
exists for all $R >0$.  At $R = 0$, this distribution becomes non-normalizable
and there is a transition to a phase with broken
ergodicity\cite{BOUCHAUD,ZINN}.  As this glass transition is approached, the relaxation
within the traps leads to the appearance of a regime in frequency space, $R << \omega \leq
a/R$, where the response is essentially frequency independent. The parameter $a$ 
defines how fast $R$ approaches zero as the glass transition temperature 
is approached.
At frequencies
smaller than a cutoff, ${\omega}_c \rightarrow 0$ as $R \rightarrow 0$, the
response is described by a power law with a positive exponent.  The exact
relationship between ${\omega}_c$ and $R$ and the exponent of the power law
depend on specific characteristics of the model.  In particular, we find that
if there is  no
correlation between the depth of a trap and its position inside the
megavalley, the peak approaches zero algebraically and is accompanied by a
flattening out of the {\it low-frequency} response.  This is reminiscent of the
observations in spin glasses\cite{NAGELSG}.  If, on the other hand, correlations 
are
introduced such that the deeper subwells are positioned further away from the
center of the megavalley, then the shift of the peak follows a
Vogel-Fulcher type law and there is no evolution of the low-frequency power law: a
scenario that is reminiscent of the behavior of supercooled
liquids\cite{NAGEL}.

Fig 1 shows a set of relaxation spectra for the model with
no correlations. It demonstrates  clearly that the low frequency power law approaches
zero as the curvature is reduced and that there is a high-frequency regime,
beyond the peak
which 
grows and becomes flatter as the curvature is reduced. Whether or not the
high-frequency cutoff 
to the power law ($\omega \simeq a/R$) is visible within the same window as the
peak ($\omega \simeq R$) depends on the parameter $a$. 
The inset shows that
there are three distinct frequency regimes and these have a striking resemblance
to observations in spin glasses\cite{NAGELSG}.  


Each trap, or valley,  is represented by a subwell and 
characterized not only by the time it takes to escape from it but also by its
internal relaxation time. 
All subwells in the free-energy as well
as the overall megavalley are assumed to be parabolic ({\it cf} Fig. 2). 
Each subwell is parameterized by its curvature $r_n$, width
$\D_n$, position  of the center $\pn$ and position of the minimum $C_n$. The set
of $\{C_n\} $ is fixed by the requirement that free energy $F(\p)$ is a
continuous function.  To simplify the picture even further, we set all $\D_n =
\D$ which  then automatically fixes $\pn = n\D$.  The curvatures are taken to 
be independent random variables picked from a distribution $P(r,n)$.
This defines a free-energy functional;
\begin{equation}
\label{eq_F}
F(\p) = {1\over 2}R\p^2 + {1\over 2}\sum_{n=-\infty}^{\infty}
\mu_n\Bigl\{r_n(\p-\pn)^2 + C_n\Bigr\},
\end{equation}    
where $\mu_n$ is the support of the $n$-th subwell. 

The dynamics is modeled by  relaxation in this free-energy surface and is
defined by 
the  Langevin equation
\begin{equation}
\label{eq_Lang}
{\pa\p\over\pa t} = - R\p - \sum \mu_nr_n(\p-\pn) + \eta(t),
\end{equation} 
where $\eta$ is a Gaussian noise with zero average and variance  
$<\eta(t)\eta(t')>=\G\d(t-t')$. The temperature scale is set by $\b=\D^2/\G$ and 
it 
is useful to introduce the ``effective'' curvature of a subwell;
$R_n=R+r_n$. Specific features of the  distribution $P(r,n)$  affect the
detailed nature of the response.
The assumption that
the curvature
of each  subwell is uncorrelated with its position, $P(r,n)\sim P(r)$ is the
easiest to implement and is the scenario that we examine in detail.  The
changes occurring due to correlations are then discussed in the light of these 
calculations.  

A natural candidate for $P(r)$ is an exponential distribution
$P(r)=e^{-\b_0r}/\b_0$.  This type of distribution has been observed in many
spin-glass models\cite{BOUCHAUD}, where it leads to power law
distributions  of
escape times.  Such distributions have also been observed in
some Ising-like systems  (without quenched
disorder) which
demonstrate glassy  behavior\cite{LEIGU}.  It has been
shown that, in
absence of the  overall curvature, $\b=\b_0$ is the temperature where the system
falls out of equilibrium\cite{BOUCHAUD}. Our model has two characteristic
temperature scales, one set by $\b_0$ and the other by $R$.

Although  Eq.(\ref{eq_Lang}) is piecewise linear inside
each subwell,  the long time evolution of the system is a non-uniformly biased
random walk  in a random environment and there is no exact solution to this
problem. However, the existence of an equilibrium distribution can be proven
rigorously.  The equilibrium probability distribution is given by $\exp (-\b
F(\p))$ as long as the integral of this function over $\p$ remains finite\cite{ZINN}. A
simple calculation shows that this integral diverges as $1/{(\b_0 -\b){\sqrt
R}}$.  
In the absence of the overall curvature the $1/{\sqrt R}$ factor gets replaced
by the system size and the probability density diverges as $\b \to
\b_0$\cite{BOUCHAUD}.
As $R \to 0$, the system falls out of equilibrium and there is a
change from exponentially decaying correlations to power-law correlations in time.
This transition 
can be studied by analyzing the {\it equilibrium}
correlation functions and the response functions associated with these through
the fluctuation dissipation relation\cite{Goldenfeld}.

A non-zero overall curvature and the requirement that the subwells
match at their boundaries, lead to a dependence of the effective barrier heights 
on the position and curvature of the subwells ({\it cf} inset of Fig. 2).  The 
wells become asymmetric and
the lower barrier of the $n$-th well is given by
\begin{equation}
\label{eq_Trap}
\D F_n = {R\over 2\r}({1\over 2}-n\r)^2,
\end{equation} 
where $\r=R/R_n$. This expression is true for $n\r<1/2$, otherwise $\D
F=0$.  Clearly, the
dynamics in  subwells $\D F>0$ and those with $\D F=0$ is very different. In
fact, it is  possible to talk about internal dynamics only when the internal
relaxation  time $(r_n)^{-1}$  is shorter than the escape time, which is $\sim
{\b}\exp(\b\D  F_n(r))$. This suggests writing the correlation function, $C(t)$, as
a sum of  two components: internal relaxation and barrier crossing contributions
from the  wells with non-zero barriers and a ``free relaxation'' inside the
megavalley  coming from the subwells where the barriers have vanished due to the
overall  curvature. The weakest point in our analysis is  the treatment of 
the free relaxation.  In the absence of an exact solution, we have described
this as simply a Debye response weighted by the effective number of barrier-less 
wells:
\begin{eqnarray}
\label{eq_C1}
C(t)& =& C(t)^{trap}+C(t)^{free} \\ \nonumber
C(t)^{trap} &= &\sum_n\theta({1\over 2}-n\r) e^{-\b F_n^{min}}
	\Bigl({e^{-R_nt}\over R_n} + e^{-{t\over\b}e^{-\b\D F_n}}\Bigr)
\\ \nonumber 
C(t)^{free}& = &\sum_n\theta(n\r-{1\over 2}){e^{-\b Rn^2-Rt}\over R}.
\end{eqnarray}
In writing these expressions, we have used that fact that an equilibrium
distribution exists and the probability
of finding the system  in subwell $n$ is given by the Boltzmann factor,
$\exp({-\b F_n^{min}})$, where $F_n^{min}$ is the minimum value of the free energy 
inside a subwell.
Assuming that the local 
curvature and the 
position are uncorrelated, and estimating the contribution of the trapped and free 
relaxation from the ``average'' weights of
barrier and  barrier-less wells at each position $n$, leads to:
\begin{equation}
\label{eq_Av_Trap}
\bar C(t)^{trap} = \sum_n\int_{n\r<1/2}{dr} P(r)e^{-\b F_n^{min}}
	\Bigl({e^{-R_nt}\over R_n} +e^{-{t\over\b}e^{-\b\D F_n}}\Bigr),
\end{equation}
where the first term describes the internal relaxation within subwells and the
second term describes activated processes. The free part can be written in a
similar fashion.

One of the most interesting aspects of our model is the high-frequency response
and the origin of this can be understood from an analysis of the 
limiting model where $R =  0$ and $\beta \rightarrow
{\beta}_0$\cite{BOUCHAUD}. In this limit, the hopping
term in  $C^{trap}$ becomes identical to the one analyzed in \cite{BOUCHAUD}. In
frequency  domain, the imaginary part of the susceptibility, $\chi ^{''}(\o) = \o C(\o)$,
is known  to grow as  $\o^{(\b_0-\b)/\b}$ at small frequencies and decays at
large $\o$  as $1/\o$. However, there is a new feature arising  from the internal
dynamics  which drastically changes the high frequency behavior.
The internal relaxation part of $\bar C^{trap}$ is given by
\begin{equation}
\label{eq_trap_int}
\bar C(\o)^{trap,int}=\int_{r^*}^{\infty}{dr}{e^{-(\b_0-\b)r}\over \o^2+r^2}.
\end{equation}
As $\b\to\b_0$, the contribution of this term to $\chi (\o)$ becomes $({\pi}/2 -
{\rm atan}(r^*/{\o}))$; an extremely slowly decaying function of frequency which
has a high-frequency cutoff $\simeq 1/(\b - \b_0)$\cite{foot1}.  
The total frequency-dependent response, therefore, behaves as
$\o^{(\b_0-\b)/\b}$ for $\o\to 0$ and is a slowly-decaying function at large
frequencies. The free (Debye) part is not present in this model with 
$R = 0$.

The overall curvature 
changes the effective barriers heights and therefore influences the response at
both high and low frequencies.  To simplify the analysis, we adopt a specific
relationship  between $R$ and $\b_0$, and 
choose the temperature at which 
$R=0$ to be at  $\b_0$ in such a way that $a(\b_0-\b) = \b_0\b R$ and take $a = 1$.  
The equilibrium relaxation spectra are then given by,
\begin{equation}
\label{C_long}
C^{trap}(\omega) = \sum_ne^{-\beta Rn^2}\int_{nR}^{\infty}{dr}
	 e^{-(\beta_0-\beta) r}
	\Bigl({1\over\omega^2+ r^2} 
                 +{e^{-\beta\Delta F(r,n)}/\beta\over \omega^2+ 
                 e^{-2\beta\Delta F(r,n)}/\beta^2}  \Bigr).
\end{equation} 

In this expression, the first term arises from the internal relaxation, and is
identical to  (\ref{eq_trap_int}), except for the cutoff being determined
by $R$.  The second term is the ``hopping'' contribution, and can be 
rewritten,  after a change
of  variables, as
\begin{equation}
\label{eq_trap_hop}
\bar C(\o)^{trap,hop}={1\over\beta^2}\sum_ne^{-\beta Rn^2}\int^{\exp(-\beta nR)}_{0}{dz}
                {z^{-1+\beta_0/\beta}\over
		\omega^2+ z^2/\beta^2}. 
\end{equation}  

The overall curvature enters through the upper cutoff in the integral, and as
$R\to 0$, exhibits the slow dynamics discussed in the previous paragraph. There
is a tradeoff between the hopping dynamics and free relaxation with the hopping
contribution  decreasing as
$R$ increases and more and more wells become effectively ``free''.
The free part is
given by, 
\begin{equation}
\label{eq_free}
\bar C^{free}(\o) = {1\over R^2+\o^2}\sum_ne^{-\beta Rn^2} \int^{nR}_{0}{dr}e^{-\b_0r},
\end{equation}  
and leads to a Debye spectrum with peak at $\o=R$.  Numerical results
based on
this description of the response are shown in Fig. 1.  The contribution from the 
deep traps shows one essential characteristic of the model; the change in the
high-frequency response accompanying the low-frequency changes as $R \to 0$.  The 
inset demonstrates that the free part is responsible for the
shift in peak frequencies and the appearance of an intermediate regime.

There are three
different  regimes of the response: low frequency power law, associated
with  hopping between different subwells; Debye-like peak coming from
barrier-less  relaxation and high frequency power law decay as a result of
superposition of  multiple single relaxation time processes. The weakest point
in this  construction is the intermediate, Debye contribution which in principle
overlaps  with both low and high frequency tails. This crossover region we
expect to be  very sensitive to all the approximations made. On the other hand,
the low-frequency and high-frequency regimes are relatively insensitive to these 
approximations 
since they depend on processes related to deep
subwells. 

This picture is very similar to the experimentally observed response  
in  spin glasses \cite{NAGELSG}, with the response flattening out at both high
and low frequency ends and the peak moving much slower than exponentially. In supercooled
liquids, only the high frequency piece flattens out, and 
the  peak shifts towards zero frequency according to a Vogel-Fulcher
law\cite{REVIEW,NAGEL}.  From the viewpoint of our model this
difference can be ascribed to a difference in the nature of correlations in the two
systems. In a structural glass, the crystalline state is the absolute global
free-energy minimum. The subwells of our model correspond to metastable states
and the megavalley represents the states accessible in the supercooled phase,
with the crystalline minimum lying outside this region.
This suggests a model where the depth of a subwell and its position are correlated and
the deeper wells are situated further away from the minimum of the megavalley.
A correlation in $P(r,n)$, of the form suggested above, leads to an
upper cutoff in the distribution of barrier heights ({\it cf}
Eq. (\ref{C_long})).
This cutoff leads to a maximum escape time
$t_{esc}  \sim
\exp({1/R}^\alpha)$ where $\alpha >0$ is the exponent of a power law describing   the
correlation between the depth and the position of subwells. Beyond this time
scale,  there is only
barrier-free motion in our model and, therefore, at frequencies
lower than ${\o}_{c} = (t_{esc})^{-1}$, we predict that $\chi(\o) \sim \o$
and  there is
no flattening out of the low-frequency response. The upper cutoff does not have
a large  influence on the
response arising from the internal dynamics of the subwells.  The high-frequency 
cutoff is still approximately given by $1/R$ as can be easily seen from 
Eq. (\ref{eq_trap_int}). 

In conclusion, we have demonstrated that the basic features of the relaxation
spectra near the glass transition can
be understood on the basis of a multivalleyed free-energy surface with an
overall curvature which goes to zero at the glass transition. The spectrum
crosses over from being pure Debye at large curvatures to one with three distinct
regimes with an asymptotic high-frequency power law characterized by an exponent 
approaching zero as the curvature goes to zero.  The model is also characterized 
by the distribution of curvatures of the subwells.  Our analysis suggests that 
the relaxation
spectra of spin glasses
and structural glasses can be described by the same underlying model with
different correlations in the distribution of subwells.

The original motivation for constructing this model came
from observations in a frustrated spin model without quenched
disorder\cite{LEIGU} whose phenomenology is remarkably similar to structural
glasses.   Simulations of this model indicated a free-energy surface 
with an overall curvature and the vanishing of this curvature was accompanied by 
the appearance of broken ergodicity and  ``aging''\cite{LEIGU}.  This same model 
also showed a power-law distribution of trapping times in the
subwells\cite{LEIGU}. This spin model can, therefore, be viewed as a
microscopic realization of the dynamical model presented here. Since it can also be
viewed as a model for real glasses, further
studies of this model should lead to an understanding of the connection between
spatial inhomogeneities and the anomalous relaxations near a glass transition
and clarify the relevance of our dynamical model to  real glasses.

\stars

The authors would like to acknowledge the hospitality of ITP, Santa Barbara
where a major portion of this work was performed.  This work has been partially
supported by NSF-DMR-9520923.

\begin{figure}[h]
\epsfxsize=7.0 cm \epsfysize=3.0in
\epsfbox{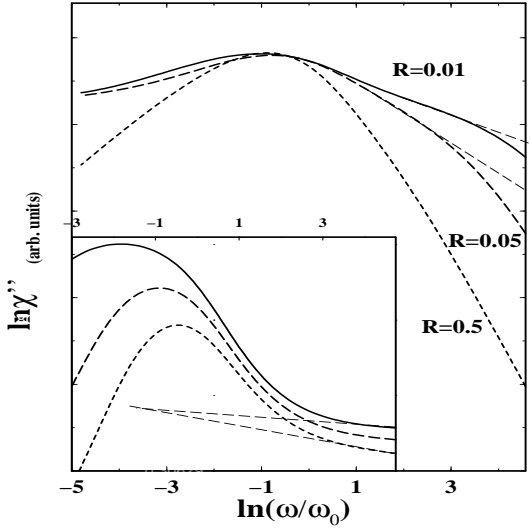}	
\caption{Contribution of deep traps to the imaginary part of susceptibility,
$\chi ^{''} (\o)$, for the
model with no correlations, calculated 
using Eq. (\ref{C_long}), for different values of $R$. The peak intensities have 
been matched. $\omega_{0}$ is a
microscopic frequency scale determined by the parameters of the model. 
The inset shows the full
relaxation spectra, {\it including} the contribution from the ``free'' subwells for
$R=$ 0.1, 0.05 and 0.01.  It is seen 
that the free part is responsible for the shift of the peak
and the appearance of an intermediate regime. The long-dashed  lines show the 
high-frequency power law extending up to $\omega \simeq a/R$. The parameter
$a=1$ in the main figure but $a=100$ in the inset.}
\label{fig1}	
\end{figure}	

\begin{figure}[h]
\epsfxsize=7.0 cm \epsfysize=3.0in
\epsfbox{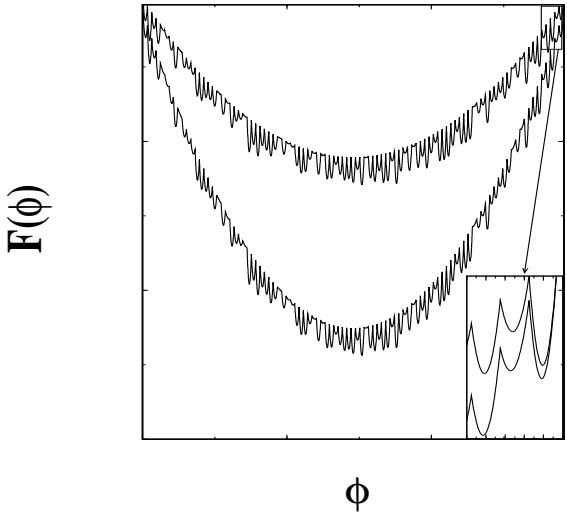}	
\caption{Free energy landscape for two different values of $R$ with a fixed
distribution of  $\{r_n\}$. The inset shows how the overall curvature modifies
the heights of  the barriers between the subwells.}
\label{fig2}	
\end{figure}

\end{document}

%% file: euromacr.tex
\def\stars{\bigskip\centerline{***}\medskip}

\newif\ifboo \boofalse


%% file: euro.bbl
\begin{thebibliography}{99}
\bibitem{foot1}The lower cutoff $r^*$ is introduced to ensure that one can define 
an internal relaxation process.  This arbitrary cutoff is necessary only for $R 
=0$ and does not affect the high-frequency behavior.
\end{thebibliography}
